\documentclass[a4paper,twocolumn,superscriptaddress,11pt,accepted=2019-06-04]{quantumarticle}
\pdfoutput=1
\usepackage[utf8]{inputenc}
\usepackage[english]{babel}
\usepackage[T1]{fontenc}
\usepackage{amsmath}
\usepackage{hyperref}

\usepackage{tikz}
\usepackage{lipsum}

\usepackage[numbers,sort&compress]{natbib}

\usepackage{amsmath, amssymb}	
\usepackage{bbm}
\usepackage{bm}				
\usepackage{braket}				
\usepackage{graphicx} 			
\usepackage{mathtools}
\usepackage{cleveref}
\usepackage{colonequals} 
\newcommand{\ce}{\colonequals}

\usepackage{hyperref}

\renewcommand{\Im}{\mathop{\rm Im}}	

\newcommand{\kket}[1]{\left| \left. #1 \right> \right>} 		
\newcommand{\bbra}[1]{\left< \left<  #1 \right. \right|} 	

\newcommand{\nn}{\nonumber}		

\usepackage[numbers,sort&compress]{natbib}
 
\begin{document}

\title{Quantizing time: Interacting clocks and systems}
\date{\today}
\author{Alexander R. H. Smith}
\email[]{alexander.r.smith@dartmouth.edu}
\affiliation{Department of Physics and Astronomy, Dartmouth College, Hanover, New Hampshire 03755, USA}
\orcid{0000-0002-4618-4832}

\author{Mehdi Ahmadi}
\email[]{mahmadi@scu.edu}
\affiliation{Department of Mathematics and Computer Science, Santa Clara University, Santa Clara, California 95053, USA}
\orcid{0000-0001-6646-413X}

\maketitle

\begin{abstract}
This article generalizes the conditional probability interpretation of time in which time evolution is realized through entanglement between a clock and a system of interest. This formalism is based upon conditioning a solution to the Wheeler-DeWitt equation on a subsystem of the Universe, serving as a clock, being in a state corresponding to a time $t$. Doing so assigns a conditional state to the rest of the Universe $\ket{\psi_S(t)}$, referred to as the system. We demonstrate that when the total Hamiltonian appearing in the Wheeler-DeWitt equation contains an interaction term coupling the clock and system, the conditional state $\ket{\psi_S(t)}$ satisfies a time-nonlocal Schr\"{o}dinger equation in which the system Hamiltonian is replaced with a self-adjoint integral operator. This time-nonlocal Schr\"{o}dinger equation is solved perturbatively and three examples of clock-system interactions are examined. One example considered supposes that the clock and system interact via Newtonian gravity, which leads to the system's Hamiltonian developing corrections on the order of $G/c^4$ and inversely proportional to the distance between the clock and system.
\end{abstract}

\section{Introduction}
\label{Introduction}

In quantum theory, time enters through its appearance as a classical parameter in the Schr\"{o}dinger equation, as opposed to other physical quantities, such as position or momentum, which are associated with self-adjoint operators and treated dynamically. Operationally, this notion of time is what is measured by the clock on the wall of an experimenter's laboratory. The clock is considered to be a large classical object, not subject to quantum fluctuations, and does not interact with the system whose evolution it is tracking. Quantum theory describes the evolution of physical systems with respect to such a clock.

However, the canonical quantization of gravity leads to the Wheeler-DeWitt equation: physical states are annihilated by the Hamiltonian of the theory. In other words, the wave function of the Universe\,---\,which includes the experimenter's clock, the system the experimenter is interested in, and everything else\,---\,is in {an eigenstate} of its Hamiltonian. Combined with the Schr\"{o}dinger equation, the Wheeler-DeWitt equation dictates that the physical states of the theory do not evolve in time {(i.e., the Hamiltonian of the theory does not generate time translations of the physical states with respect to an external time)}. How then do we explain the time evolution we see around us? This dilemma constitutes one aspect of the problem of time, and is sometimes phrased as the problem of finding a background-independent quantum theory~\cite{Isham1993, Kuchar:2011}.

A necessary requirement for any quantum theory of gravity is to answer this question and explain how the familiar Schr\"{o}dinger equation comes about from the Wheeler-DeWitt equation. The conditional probability interpretation of time offers an answer. As introduced by {Page and Wootters} \cite{Page:1983, Wootters:1984, Page:1989br, Page1994}, the conditional probability interpretation defines the state of a system at a time~$t$ as a solution to the Wheeler-DeWitt equation conditioned on a subsystem of the Universe, serving as a clock, to be in a state corresponding to the time $t$. Given an appropriate choice for the Hamiltonian of the Universe and choice of clock, one finds this conditional state of the system satisfies the Schr\"{o}dinger equation.

This interpretation of time was initially criticized by Kucha\v{r}~\cite{Page1994, Kuchar:2011}, who argued that it was unable to reproduce the correct two-time correlation functions, i.e., supposing a system was initially prepared in some state, what is the probability of finding the system in a different state at a later time? This criticism has since been overcome in two different ways: first, by correctly formulating the two-time correlation functions in terms of physical observables \cite{Dolby2004}, and second, by modelling the measurement of the two-time correlation function as two successive von Neumann measurements~\cite{Giovannetti:2015}.

Recently, the conditional probability interpretation of time has received considerable attention. Gambini  {\emph{et al.} \cite{Gambini:2004b,Gambini:2004, Gambini:2009}} have demonstrated that the conditional probability interpretation can result in a fundamental decoherence mechanism
and explored the consequences of this fact in relation to the black hole information loss problem~\cite{Gambini:2004a}. Leon and Maccone~\cite{Leon:2017} have shown that this interpretation overcomes Pauli's objection to constructing a time operator in quantum mechanics. More recently, the the conditional probability interpretation of time has been studied in the context of  quantum reference frames and quantum resource theory of coherence \cite{Martinelli:2019, Mendes:2018}. Furthermore, conditional probabilities were employed to illustrate a novel
quantum time dilation effect induced by nonclassical states of relativistic particles whose internal degrees of freedom serve as clocks~\cite{Smith:2019}. Others have applied the formalism to a number of different systems and commented on various aspects of the proposal~\mbox{\cite{Corbin:2009, Ekaterina-Moreva:2014, Marletto:2016, Boette:2016, Bryan:2017}}.

The purpose of this article is to extend the conditional probability interpretation of time to take into account the possibility that the system being employed as a clock interacts with a system whose evolution the clock is tracking. As gravity couples everything, including clocks, this extension is necessary if the conditional probability interpretation of time is to be applied in a quantum gravitational setting. We find that taking into account a possible clock-system interaction within the conditional probability interpretation of time results in a time-nonlocal modification to the Schr\"{o}dinger equation.

We begin in Sec.~\ref{The Hamiltonian constraint in classical and quantum mechanics} by reviewing the timeless formulation of classical mechanics \cite{Rovelli:2004, Kiefer:2012} and its quantization. In Sec.~\ref{The conditional probability interpretation of time} we introduce the conditional probability interpretation of time and its generalization to interacting clocks and systems. This results in a modified time-nonlocal Schr\"{o}dinger equation, for which we derive a series solution. {In Sec.~\ref{examples} we give three explicit examples of clock-system interactions: the first leads to a time-dependent modification to the system Hamiltonian, the second is of a clock and system interacting gravitationally, and the third examines the case when the clock and system are two-level systems.} We conclude in Sec.~\ref{ch8Summary} with a summary of our results and comment on future directions of research.

\begin{figure}[t]
\includegraphics[width = .48 \textwidth]{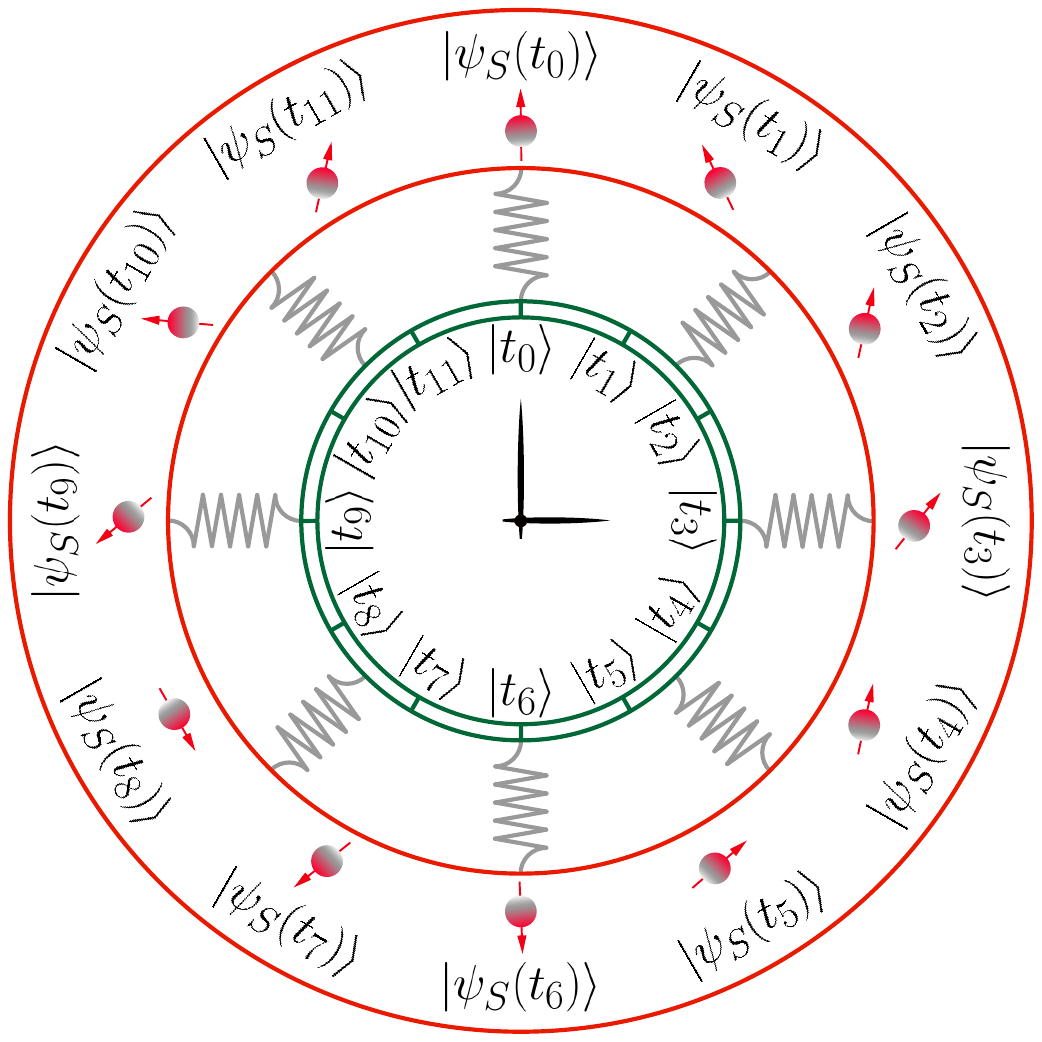}
\caption{A clock depicted in green is used to track the evolution of a system depicted in red. The conditional probability interpretation is extended to allow for the clock and system to interact, this interaction is depicted by the grey springs.
}
\label{time}
\end{figure}



\section{The Hamiltonian constraint in classical and quantum mechanics}
\label{The Hamiltonian constraint in classical and quantum mechanics}

Both classical and quantum mechanics describe the evolution of the state, or equivalently the observables, of a physical system in time. However, general relativity demands that time itself be treated like any other physical system, that is, time should be treated dynamically. In this context, the dynamics of both classical and quantum mechanics describe relations between two physical systems: a clock, which indicates the time, and everything else. In this section we present a formulation of classical mechanics in which time is treated dynamically and on equal footing with the system whose evolution we are interested in. We then pass over to the corresponding quantum theory \`{a} la Dirac \cite{Dirac:1964}.

Consider a system $S$ described by the action
\begin{align}
\mathcal{S} = \int_{t_1}^{t_2} dt \, L_S \!\left(q,q'\right), \nn
\end{align}
where $L_S\! \left(q,q'\right)$ is the Lagrangian associated with $S$, $q=q(t)$ denotes a set of generalized coordinates describing $S$, and $q'=q'(t)$ denotes the differentiation of these coordinates with respect to $t$.

Let us introduce an integration parameter $\tau$ and promote $t$ to a dynamical variable $t(\tau)$, which we associate with the reading of a clock~$C$. Through application of the chain rule, the action may be expressed as
\begin{align}
\mathcal{S} = \int_{\tau_1}^{\tau_2} d\tau \, \dot{t} L_S\! \left(q,\dot{q}/\dot{t}\right) = \int_{\tau_1}^{\tau_2} d\tau \,  {L}\!\left(q,\dot{q}, \dot{t}\right), \nn 
\end{align}
where ${L}\!\left(q,\dot{q}, \dot{t}\right) \ce \dot{t} L_S\!\left(q,\dot{q}/\dot{t}\right)$ is the Lagrangian describing both $C$ and $S$ and the dot denotes differentiation with respect to $\tau$.

The Hamiltonian associated with ${L}\left(q,\dot{q},\dot{t}\right)$ is obtained by a Legendre transformation with respect to both $\dot{q}$ and~$\dot{t}$
\begin{align}
\tilde{H} &= p_t \dot{t} + p_q \dot{q} - L(q,\dot{q},\dot{t}) = \dot{t} \left( p_t + H_S \right), \label{Htilde}
\end{align}
where $H_S \ce p_q q' - L_S \!\left(q,q'\right)$ is the Hamiltonian associated with $L_S \!\left(q,q'\right)$ and we have used the fact that the momentum conjugate to $q$ defined by ${L}\left(q,\dot{q},\dot{t}\right)$ is
\begin{align}
p_q \ce \frac{\partial L \!\left(q,\dot{q},\dot{t}\right)}{\partial \dot{q} } =  \dot{t} \frac{\partial L_S \!\left(q,\dot{q}/ \dot{t}\right)}{\partial \left(\dot{q} / \dot{t} \right) } \frac{1}{\dot{t}} =  \frac{\partial L_S\left(q, q'\right)}{\partial q' }, \nn
\end{align}
which coincides with the momentum conjugate to $q$ defined by $L_S\left(q,q'\right)$. The momentum conjugate to $t$ is
\begin{align}
p_t \ce \frac{\partial L\!\left(q,\dot{q},\dot{t}\right)}{\partial \dot{t} } =  L_S\!\left(q, q' \right)  - q' p_q =- H_S. \label{pt}
\end{align}
In light of Eq.~\eqref{pt}, we see that the term inside the brackets in Eq.~\eqref{Htilde} is constrained to vanish
\begin{align}
H \ce p_t + H_S \approx 0, \label{totalHamiltonian}
\end{align}
where $\approx$ indicates weak equality in the sense of Dirac~\cite{Dirac:1964}. We will refer to $H$ as the total Hamiltonian (a.k.a. the super Hamiltonian or Hamiltonian constraint) as it describes both $C$ and $S$.

It is natural to ask if the total Hamiltonian given in Eq.~\eqref{totalHamiltonian} is the most general possible. The answer is no. The total Hamiltonian can differ in two important ways:
\begin{enumerate}
\item An additional term $H_{int} = H_{int}(t, p_t, q,p_q)$ may be included in the total Hamiltonian, which couples $C$ and $S$; this term will be referred to as the interaction Hamiltonian.

\item The momentum $p_t$ may be replaced by a function of the conjugate variables associated with $C$, which we will refer to as the clock Hamiltonian and denote by $H_C = H_C(t, p_t)$. Note that in general  $C$ may be a composite system and have more than just one pair of conjugate variables.
\end{enumerate}
Accounting for these generalizations, the most general total Hamiltonian is
\begin{align}
H = H_C + H_S +  H_{int} \approx 0. \label{GeneralTotalHamiltonian}
\end{align}

Motivating these generalizations is our best theory of time: general relativity. The Hamiltonian formulation of general relativity does not admit a total Hamiltonian of the form given in Eq.~\eqref{totalHamiltonian}. Total Hamiltonians that are linear in one of the conjugate momenta, like Eq.~\eqref{totalHamiltonian}, indicate there is a preferred time variable in the theory~\cite{Dirac:1933}, and this structure is not present in general relativity. Further, gravity couples everything, including a clock and the system whose evolution it is tracking. Therefore, in a gravitational setting, we should expect an interaction Hamiltonian $H_{int}$ to appear in the total Hamiltonian $H$ coupling $C$ and $S$. We note that the inclusion of such an interaction Hamiltonian may result in a total Hamiltonian that is nonlinear in the conjugate momentum associated with the clock, even if the clock Hamiltonian is proportional to this momentum.

We now wish to quantize the theory described by the total Hamiltonian given in Eq.~\eqref{GeneralTotalHamiltonian}. To do so, we  follow the prescription given by Dirac \cite{Dirac:1964}.  We associate with  $C$ and  $S$ the Hilbert spaces $\mathcal{H}_C$ and $\mathcal{H}_S$, respectively. The total Hamiltonian $H$ becomes an operator acting on the kinematical Hilbert space $\mathcal{H}_{\rm kin}\simeq\mathcal{H}_C\otimes\mathcal{H}_S$, and the constraint in Eq.~\eqref{GeneralTotalHamiltonian} becomes
\begin{align}
H \kket{\Psi} &= \big(H_C \otimes I_S + I_C \otimes H_S +  H_{int} \big) \kket{\Psi}\nn \\
& =0 \label{constraint},
\end{align}
where $I_C$ and $I_S$ denote the identity operators on $\mathcal{H}_C$ and $\mathcal{H}_S$, respectively. {Within the context of quantum gravity, Eq.~\eqref{constraint} is referred to as the Wheeler-DeWitt equation.} The double ket notation is used to remind us that $\kket{\Psi}$ is a state of both the clock and system. States $\kket{\Psi}$ satisfying the constraint are in the physical Hilbert space $\mathcal{H}_{\rm phy}$; states in the physical Hilbert space $\kket{\Psi} \in \mathcal{H}_{\rm phy}$ will be referred to as physical states. To completely specify the physical Hilbert space~$\mathcal{H}_{\rm phy}$ one must also choose an inner product on $\mathcal{H}_{\rm phy}$, which we will do in the following section.

In general, the physical states evolve unitarily with respect to an external time, this evolution being generated by the total Hamiltonian. However, in totally constrained theories, such as the one defined by Eq.~\eqref{constraint}, the physical states are annihilated by the total Hamiltonian, \mbox{$H \kket{\Psi} =0$}, and therefore do not evolve with respect to any external time. The question then arises, how do we recover the dynamics we see around us from the frozen state $\kket{\Psi}$? How does the Schr\"{o}dinger equation come about from the constraint $H \kket{\Psi} =0$?

These questions constitute one aspect of the problem of time in quantum gravity \cite{Isham1993, Kuchar:2011}. The conditional probability interpretation of time offers a way to reconcile the fact that the physical states are frozen with the time evolution  described by quantum theory. In this formalism the dynamics of $S$ are encoded in entanglement shared between $C$ and $S$ in the physical state~$\kket{\Psi}$. The outcome of a measurement of an observable on $S$ at a specific time~$t$, is interpreted as a measurement of the physical state $\kket{\Psi}$ conditioned on the clock being in a state corresponding to the time $t$.


\section{The conditional probability interpretation}
\label{The conditional probability interpretation of time}

We now introduce the conditional probability interpretation of time for theories described by the general total Hamiltonian given in Eq.~\eqref{constraint}{, which includes the possibility of an interaction Hamiltonian $H_{int}$ coupling $C$ and~$S$}. We will introduce the state of the system at time~$t$ by conditioning a solution to the constraint $\kket{\Psi}$ on the clock being in a state corresponding to the time~$t$. This state of the system will be seen to satisfy the Schr\"{o}dinger equation in the limit where $H_{int}$ vanishes.

\subsection{The modified Schr\"{o}dinger equation}
\label{TheModifiedSchrodingerEquation}

In the classical theory specified by the total Hamiltonian given in Eq.~\eqref{totalHamiltonian}, time is defined operationally as the outcome of a measurement of the phase space variable~$t$ associated with a clock governed by the Hamiltonian  $H_C = p_t$. In this case, the variable $t$ is canonically  conjugate to the clock Hamiltonian.

The quantized version of this notion of time is to define time as a measurement of a time operator~$T$ on the clock Hilbert space $\mathcal{H}_C \simeq L^2(\mathbb{R})$ which is canonically conjugate to the clock Hamiltonian $H_C$, ($\hbar=1$)
\begin{align}
[T, H_C] = i. \label{ClockCCR}
\end{align}
In other words, states of the clock indicating different times correspond to eigenstates $\ket{t}$ of the time operator~$T$, and the associated eigenvalue $t$ is the time indicated by the clock. Employing the Baker-Campbell-Hausdorff formula, we see that as a consequence of the commutation relation in Eq.~\eqref{ClockCCR}, $H_C$ generates translations of $T$
\begin{align}
e^{-iH_C s} T e^{iH_C s} = T - s I_C. \label{BCHtranslations1}
\end{align}
Resolving the identity on $\mathcal{H}_C$ as $I_C = \int dt \, \ket{t}\!\bra{t}$ and making use of the spectral representation of the time operator $T = \int dt \, t \ket{t}\!\bra{t}$, Eq.~\eqref{BCHtranslations1} may be expressed as
\begin{align}
\int dt \, t e^{-iH_C s}   \ket{t}\!\bra{t} e^{iH_C s} &= \int dt \, (t - s )\ket{t}\!\bra{t}\nn\\ &= \int dt \, t \ket{t+s}\!\bra{t+s}, \nn
\end{align}
or
\begin{align}
e^{-iH_C s} \ket{t} = \ket{t+s}, \label{ClockStates}
\end{align}
up to an overall phase.

{Following Page and Wootters~\cite{Page:1983, Wootters:1984}, we will take Eq.~\eqref{ClockStates} to define the states of the clock corresponding to a particular instant of time $t$, that is, $\ket{t}\ce e^{-iH_Ct} \ket{t_0}$ where $\ket{t_0} \in \mathcal{H}_C$; all other states of the clock are not to be associated with a definite value of $t$. We refer to the set of states $\{ \ket{t} \, | \, \forall \, t\}$ as the clock states. In what follows, we will also demand that the clock states form a resolution of the identity $I_C \propto \int dt \ket{t}\!\bra{t}$, which ensures that they define a normalized positive operator valued measure. The translation property expressed in Eq.~\eqref{ClockStates} insures that this POVM is covariant with respect to the clock Hamiltonian $H_C$~\cite{holevoProbabilisticStatisticalAspects1982, Braunstein:1994, Busch:1995, Braunstein:1996}

This definition of $T$ does not necessarily imply that the clock states are eigenstates of a time operator $T$ canonically conjugate to the clock Hamiltonian $H_C$.  However, when this is the case, the time operator $T$ and clock Hamiltonian $H_C$ are, by the Stone-von Neumann theorem~\cite{Hall:2013}, unitarily equivalent to the position and momentum operators on the real line. In this case the clock behaves perfectly in the sense that the clock states are orthogonal, $\braket{t | t'} = \delta (t-t')$, so as they are perfectly distinguishable, and there is no chance that the clock will run backwards in time, i.e., $\braket{t'| e^{-iH_C \tilde{t}} |t} = 0$ for all $\tilde{t}>0$ if $t>t'$~\cite{Isham1993,Unruh:1989, Dolby2004}. The first two examples considered in Sec.~\ref{examples} employ such a perfect clock, while the third example does not.

We now define the state of the system at time~$t$ as a solution to the constraint in Eq.~\eqref{constraint}  conditioned on the clock being in the state $\ket{t}$
\begin{align}
\ket{\psi_S(t)} \ce  \big( \bra{t} \otimes I_S\big) \kket{\Psi}, \label{SystemState}
\end{align}
where $\ket{\psi_S(t)} \in \mathcal{H}_S$. The state $\ket{\psi_S(t)}$ should be thought of as the time-dependent state of the system in the conventional formulation of quantum mechanics. Note that with this definition of the system state we may express the physical state $\kket{\Psi}$ as
\begin{align}
\kket{\Psi} &= \left(  \int dt \, \ket{t}\!\bra{t} \otimes I_S \right) \kket{\Psi} \nn \\
&= \int dt \, \ket{t} \ket{\psi_S(t)}. \label{entangledPhysicalState}
\end{align}
This state is expressed pictorially in Fig.~\ref{time}.

As mentioned above, we need to choose an inner product for the physical Hilbert space $\mathcal{H}_{\rm phy}$. We will choose this inner product to be
\begin{align}
\braket{\braket{\Psi|\Phi}}_{\rm phy} \ce \braket{\braket{\Psi| \big( \ket{t}\!\bra{t} \otimes I_S \big) |\Phi}}, \label{innerprod}
\end{align}
for two states $\kket{\Psi}$ and $\kket{\Phi}$ in $\mathcal{H}_{\rm phy}$, and demand that the inner product is independent of the choice\footnote{We should emphasize this is a choice of inner product and normalization of the physical states, which may severely reduce the size of the physical Hilbert space. Furthermore, it may not be necessary to preserve the probabilistic interpretation of the system state. For example, the probabilistic interpretation may only be applicable in some limit, and it is the task of the physicist to explain how this limit comes about. To quote DeWitt on this point \cite{DeWitt:1999, Kiefer:2012}:
\begin{quotation}
``\ldots one learns that time and probability are phenomenological concepts.''
\end{quotation}
And Kiefer's clarification of DeWitt's statement \cite{Kiefer:2012}:
\begin{quotation}
``The reference to probability refers to the `Hilbert-space' problem, which is intimately connected with the `problem of time'. If time is absent, the notion of a probability conserved in time does not make much sense; the traditional Hilbert-space structure was designed to implement the probability interpretation, and its fate in a timeless world thus remains open.''
\end{quotation}} of $t$. We will also require that states in $\mathcal{H}_{\rm phy}$ are normalized with respect to this inner product
\begin{align}
1 &= \braket{\braket{\Psi|\Psi}}_{\rm phy} \nn\\&=  \left[ \int dt' \, \bra{t'} \bra{\psi_S(t')} \right] \ket{t}\!\bra{t} \otimes I_S \nn \\
&\quad \times \left[ \int dt'' \, \ket{t''} \ket{\psi_S(t'')} \right] \nn\\
&= \braket{\psi_S(t) | \psi_S(t)}. \label{normalization}
\end{align}
From the above equation, we see that this choice of inner product and normalization of the physical states ensures that the system state $\ket{\psi_S(t)}$ is properly normalized at all {times}. Consequently, we can maintain the usual probabilistic interpretation of  $\ket{\psi_S(t)}$.

Let us observe how $\ket{\psi_S(t)}$ changes with the parameter $t$ labeling the clock states by acting on both sides of Eq.~\eqref{SystemState} with $id/dt$:
\begin{align}
 i\frac{d}{dt}  \ket{\psi_S(t)} &= i \frac{d}{dt}  \big(\bra{t} \otimes I_S \big) \kket{\Psi} \nn \\
&= -\bra{t}\big( H_C\otimes I_S \big) \kket{\Psi} \nn \\
&= -\bra{t}\big( H - I_C \otimes H_S -  H_{int} \big) \kket{\Psi}. \nn 
\end{align}
\begin{figure}[t]
\includegraphics[width = .48 \textwidth]{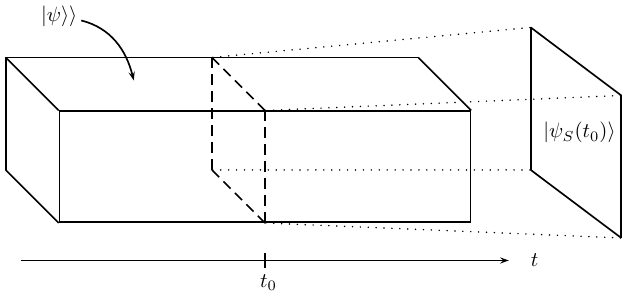}
\caption{The rectangular prism is a pictorial representation of the state $\kket{\Psi}=\int dt \, \ket{t} \ket{\psi_S(t)}$. The horizontal axis represents the Hilbert space associated with the clock $\mathcal{H}_C$ and the directions orthogonal to the horizontal axis represent the Hilbert space of the system state $\mathcal{H}_S$. The system state $\ket{\psi_S(t_0)}$ at the time $t_0$ is obtained by conditioning $\kket{\Psi}$ on the clock being in the state $\ket{t_0}$ and pictorially represented by a slice of the rectangular prism. Adapted from Giovannetti \emph{et~al.} \cite{Giovannetti:2015}.
 }
\label{time}
\end{figure}Using the fact that $H \kket{\Psi}=0$ we find $\ket{\psi_S(t)}$ satisfies
\begin{align}
i \frac{d}{dt} \ket{\psi_S(t)} = H_S \ket{\psi_S(t)} +  \bra{t} H_{int} \kket{\Psi}. \label{eqm1}
\end{align}
Inserting a resolution of the identify on $\mathcal{H}_C$ in terms of the clock states $I_C = \int dt \, \ket{t} \! \bra{t}$ between $H_{int}$ and $\kket{\Psi}$ in the second term of Eq.~\eqref{eqm1} and using the definition of the system state in Eq.~\eqref{SystemState}, we find
\begin{align}
i \frac{d}{dt} \ket{\psi_S(t)} = H_S \ket{\psi_S(t)} +  \int dt' \, K(t,t')  \ket{\psi_S(t')}, \label{eqm3}
\end{align}
where $K(t,t') \ce  \bra{t} H_{int} \ket{t'} $ is an operator acting on $\mathcal{H}_S$. We will refer to Eq.~\eqref{eqm3} as the modified Schr\"{o}dinger equation. When the interaction Hamiltonian vanishes, the modified Schr\"{o}dinger equation reduces to the usual Schr\"{o}dinger equation.

The second term on the right-hand side of Eq.~\eqref{eqm3} is a linear integral operator on $\mathcal{H}_S$ with integration kernel $K(t',t)$; we will denote this integral operator as {$H_K$ and its action as}
\begin{align}
 H_K  \ket{\psi_S(t)} &\ce  \int dt' \, K(t,t')  \ket{\psi_S(t')}. \nn
\end{align}
Note\footnote{A proof of this can be found on pages 197-198 of \cite{Yosida:1980}.}, $H_K$ is a self-adjoint operator if and only if $K(t,t')^\dagger = K(t',t)$, which is seen to be true from the definition of $K(t,t')$
\begin{align}
K(t,t')^\dagger &=  \big[ \bra{t} H_{int} \ket{t'}\big]^\dagger =  \bra{t'} H_{int} \ket{t} \nn \\
&= K(t',t), \nn
\end{align}
and therefore $H_K$ is self-adjoint.

With the definition of $H_K$, let us write the modified Schr\"{o}dinger equation in a more suggestive form
\begin{align}
i \frac{d}{dt} \ket{\psi_S(t)} = \big( H_S +  H_K  \big) \ket{\psi_S(t)}. \nn 
\end{align}
Expressed this way, the modified Schr\"{o}dinger equation can be seen as the ordinary Schr\"{o}dinger equation with the system Hamiltonian $H_S$ replaced with the self-adjoint integral operator $H_S+ H_K$.

The modified Schr\"{o}dinger equation is nonlocal in time, which means that to verify whether $\ket{\psi_S(t)}$ is a solution requires knowledge of $\ket{\psi_S(t)}$ at all times $t$, rather than a small neighbour around $t$ as is the case for the usual Schr\"{o}dinger equation. This is due to the presence of the integral operator $H_K$ whose kernel $K(t,t')$ does not vanish when $t \neq t'$.

To maintain the usual probabilistic interpretation of the system state $\ket{\psi_S(t)}$, the evolution described by the modified Schr\"{o}dinger equation must preserve its norm
\begin{align}
 \braket{\psi_S(t) | \psi_S(t)} =1,  \ \forall t \ \Rightarrow \  \frac{d}{dt} \braket{\psi_S(t) | \psi_S(t)} = 0. \nn 
\end{align}
In other words, the time evolution described by the modified Schr\"{o}dinger equation is an isometry~{\cite{Busch:1999}}. Evaluating $\frac{d}{dt} \braket{\psi_S(t) | \psi_S(t)}$ using Eq.~\eqref{eqm3} yields the condition
\begin{align}
 \int dt' \, \Im \Big[ \braket{ \psi_S(t) | K(t,t')  | \psi_S(t')}   \Big] =0. \label{restriction}
\end{align}
Since $K(t,t')\ce\braket{t| H_{int} |t'}$, Eq.~\eqref{restriction} is a condition on the interaction Hamiltonian such that the modified Schr\"{o}dinger equation preserves the norm of the system state.

We now show that Eq.~\eqref{restriction} is satisfied so long as the physical state associated with the clock and system is normalized according to Eq.~\eqref{normalization}. Using Eq.~\eqref{entangledPhysicalState} and the definition of $K(t,t')$, the left-hand side of Eq.~\eqref{restriction} can be expressed as
\begin{align}
&\bbra{\Psi} \left[  H_{int}, \ket{t}\!\bra{t} \otimes I_S \right] \kket{\Psi}\nn\\
&\qquad \qquad = \bbra{\Psi} \left[ H, \ket{t}\!\bra{t} \otimes I_S \right] \kket{\Psi} \nn \\
&\qquad  \qquad \quad - \bbra{\Psi} \left[ H_C \otimes I_S , \ket{t}\!\bra{t} \otimes I_S \right] \kket{\Psi} \nn \\
& \qquad \qquad \quad - \bbra{\Psi} \left[  I_C \otimes H_S, \ket{t}\!\bra{t} \otimes I_S \right] \kket{\Psi} \nn \\
&\qquad \qquad = - \bbra{\Psi} \big( \left[ H_C, \ket{t}\!\bra{t}\right]  \otimes I_S  \big) \kket{\Psi} \nn \\
&\qquad \qquad = -i \frac{d}{dt} \bbra{\Psi} \big( \ket{t}\!\bra{t}  \otimes I_S \big) \kket{\Psi} \nn \\
&\qquad \qquad = -i \frac{d}{dt} \braket{\braket{\Psi| \Psi}}_{\rm phy}, \nn
\end{align}
which vanishes if the physical inner product is independent of $t$. Therefore, the condition given in Eq.~\eqref{restriction} is equivalent to demanding that the physical inner product is independent of $t$, as demanded by Eq.~\eqref{normalization}.

From Eq.~\eqref{entangledPhysicalState} we see that $\kket{\Psi}$ is an entangled state of the clock and system. This entanglement encodes the time evolution of the system state $\ket{\psi_S(t)}$ generated by the modified Schr\"{o}dinger equation. This is somewhat analogous to the situation in general relativity. The state $\kket{\Psi}$ is analogous to the 4-dimensional spacetime metric\,---\,neither evolve with respect to an external time. However, one can foliate the 4-dimensional spacetime by spacelike hypersurfaces (by choosing a clock), and then the 4-dimensional metric encodes the evolution from one hypersurface to the next of the induced 3-metric and its conjugate momentum on these hypersurfaces. This evolution is analogous to the evolution of the system state $\ket{\psi_S(t)}$ governed by the modified Schr\"{o}dinger equation.

\subsection{Solving the modified Schr\"{o}dinger equation}

{To further explore the consequences of an interaction Hamiltonian $H_{int} \ce \lambda \bar{H}_{int}$, where we have made the interaction strength $\lambda\in\mathbb{R}$ explicit, we seek a series solution in $\lambda$ to the modified Schr\"{o}dinger equation.}

Suppose the modified Schr\"{o}dinger equation can be solved for $\ket{\psi_S(t)}$ in terms of a time evolution operator $V(t,t_0)$, so that the solution may be given as
\begin{align}
\ket{\psi_S(t)} = V(t,t_0) \ket{\psi_S(t_0)}, \nn 
 \end{align}
 where $\ket{\psi_S(t_0)} \in \mathcal{H}_S$ is the state of the system at the time $t=t_0$ and $V(t_0,t_0) = I_S$. Suppose $V(t,t_0)$ may be expanded in powers of $\lambda$ as
\begin{align}
 V(t,t_0) = \sum_{n=0}^\infty \lambda^n V_n(t,t_0). \nn
\end{align}
Upon substituting $\ket{\psi_S(t)} = V(t,t_0) \ket{\psi_S(t_0)}$ into the modified Schr\"{o}dinger equation and equating terms at equal order in $\lambda$, we find the operator $V_0(t,t_0)$ satisfies
\begin{align}
i \frac{d}{dt} V_0(t,t_0) &= H_S V_0(t,t_0),\nn\end{align}
which implies that
\begin{align}
V_0(t,t_0) &= e^{-iH_S (t-t_0)}, \label{0order}
\end{align}
and we see that $V_0(t,t_0)$ is the usual Schr\"{o}dinger time evolution operator $U(t,t_0) \ce e^{-iH_S (t-t_0)}$.  The higher order operators $V_n(t,t_0)$ satisfy
\begin{widetext}
 \begin{align}
i \frac{d}{dt} V_n(t,t_0) &= H_S V_n(t,t_0)  +  \int dt' \, \bar{K}(t,t')  V_{n-1}(t',t_0), \label{norder}
\end{align}
where $\bar{K}(t,t') \ce \braket{t| \bar{H}_{int} | t '}$.  The solution to Eq.~\eqref{norder}  is  given by the recurrence relation
\begin{align}
V_n(t,t_0) &=  -i   U(t,t_0) \int_{t_0}^{t} ds \, U(s,t_0)^\dagger  \int du \, \bar{K}(s,u) V_{n-1}(u,t_0). \label{nordersol}
\end{align}
Using Eqs.~\eqref{0order}  and \eqref{nordersol}, the time evolution operator $V(t,t_0)$ may be expanded to leading order in $\lambda$ as
\begin{align}
V(t,t_0)
&= U(t,t_0)\left[ I_S   -i \lambda  \int_{t_0}^{t} ds \, U(s,t_0)^\dagger  \int du \, \bar{K}(s,u) U(u,t_0) + \mathcal{O}\!\left(\lambda^2\right)\right].  \label{SeriesExpansion}
\end{align}
\end{widetext}
Equation \eqref{SeriesExpansion} is analogous to the Dyson series, and reduces to the Dyson series when the integration kernel is of the form given in Eq.~\eqref{simpAf(T)}.

{Before closing this section we note that given a solution to the modified Schr\"{o}dinger equation in terms of the time evolution operator $V(t,t_0)$, the modified Schr\"{o}dinger equation may be expressed as
\begin{align}
i \frac{d}{dt} \ket{\psi_S(t)} &= \big[ H_S + H_{mod}(t) \big] \ket{\psi_S(t)}, \nn
\end{align}
where we have defined the modified Hamiltonian 
\begin{align}
H_{mod}(t) \ce \int dt' \, K(t,t') V(t',t). \nn
\end{align}
It immediately follows from the condition expressed in Eq.~\eqref{restriction} that $H_{mod}(t)$ is self-adjoint, and thus the evolution generated by the modified Schr\"{o}dinger equation is unitary. The consequence of this is that the modified Schr\"{o}dinger equation can be recast as the usual Schr\"{o}dinger equation defined by the Hamiltonian \mbox{$H_S + H_{mod}(t)$}. Employing the series solution for $V(t,t_0)$ given above,  $H_{mod}(t)$ can be evaluated perturbatively.}

\section{Examples of clock-system interactions}
\label{examples}

In this section we examine three different clock-system interactions and examine the ensuing relational dynamics of the system that result.

\subsection{Interactions leading to time-dependent system Hamiltonians}

Consider the following interaction Hamiltonian
\begin{align}
 H_{int} =  \sum_i f_i(T) \otimes S_i, \label{specialinteraction}
\end{align}
where {$f_i(T) \ce \int dt \, f_i(t) \ket{t} \! \bra{t}$} is a self-adjoint function of the time operator $T$ satisfying Eq.~\eqref{ClockCCR} and $S_i$ is a self-adjoint operator on $\mathcal{H}_S$. The operator $K(t,t')$ associated with this interaction Hamiltonian is
\begin{align}
K(t,t') &\ce \braket{t | H_{int} | t' } \nn \\
&=  \sum_i \braket{t | f_i(T)| t'}  S_i \nn \\
&=  \sum_i \bra{t} \left(\int dt'' \,  f_i(t'') \ket{t''} \!\bra{t''} \right) \ket{t'}  S_i \nn \\
&= \delta(t-t') \sum_i f_i(t) S_i. \label{simpAf(T)}
\end{align}
One can easily verify that this $K(t,t')$ satisfies the condition in Eq.~\eqref{restriction}. Substituting Eq.~\eqref{simpAf(T)} into the modified Schr\"{o}dinger equation and simplifying, we find
\begin{align}
i \frac{d}{dt} \ket{\psi_S(t)} &=  \left[ H_S +  \sum_i f_i(t) S_i  \right]\ket{\psi_S(t)}, \nn
\end{align}
which we recognize as the ordinary Schr\"{o}dinger equation with the system Hamiltonian $H_S$ replaced with the time-dependent Hamiltonian
\begin{align}
H_S(t) = H_S +  \sum_i f_i(t) S_i . \nn
\end{align}
Therefore, an interaction Hamiltonian of the form given in Eq.~\eqref{specialinteraction} results in a time-dependent evolution of the system state.

{This example demonstrates that interactions of the form given in Eq.~\eqref{specialinteraction} result in a relational dynamics generated by a time dependent Hamiltonian $H_S(t)$. The usual interpretation of such Hamiltonians is that they arise from an external agent exerting some classical control over the system (e.g., turning a nob which increases the strength of a magnetic field). This demonstrates that clock-system couplings of the form given in Eq.~\eqref{specialinteraction} can model such time-dependent classical control. }

\subsection{{Gravitationally interacting clocks and systems}}

{Let us now consider the case in which the clock and system are coupled through Newtonian gravity as described by the interaction Hamiltonian
\begin{align}
H_{int} = - \frac{G m_C M_S}{d} = -\frac{G}{c^4 d} H_C \otimes H_S, \nn
\end{align}
where we have associated the mass of the clock $m_C$ and system $m_S$ with their respective Hamiltonians. Suppose that the clock Hamiltonian is given by its momentum operator $H_C = P_C$, as discussed at the beginning of Sec.~\ref{TheModifiedSchrodingerEquation}. In this case the integration kernel appearing in the modified Schr\"{o}dinger equation defined by the above interaction Hamiltonian evaluates to
\begin{align}
K(t,t') &\ce \braket{t| H_{int} | t'} \nn \\
 &= -\frac{G}{c^4 d} \braket{t| P_C | t'} H_S \nn \\
&= -\frac{G}{c^4 d}  \left[ \int dp \, p  \braket{t| p} \braket{p| t'} \right] H_S \nn \\
&= -\frac{G}{c^4 d}  \left[ \frac{1}{2\pi} \int dp \, p  e^{-ip( t'-t)}      \right]H_S  \nn\\
&=  -\frac{G}{c^4 d} i \delta'(t'-t) H_S. \label{gravKernal}
\end{align}
 where $\ket{p}$ are eigenkets of $P_C$ with eigenvalue $p$ and we have made use of the spectral decomposition of the clock Hamiltonian $P_C = \int dp \, p \ket{p}\!\bra{p}$ in arriving at the third equality above. }

{Substituting Eq.~\eqref{gravKernal} into the modified Schr\"{o}dinger, Eq.~\eqref{eqm3}, yields
\begin{align}
i \frac{d}{dt} \ket{\psi_S(t)} &= H_S \ket{\psi_S(t)} \nn \\
&\quad -\frac{G}{c^4 d} i \int dt' \, \delta'(t'-t) H_S \ket{\psi_S(t')} \nn\\
&= H_S \ket{\psi_S(t)}  \nn \\
&\quad + \frac{G}{c^4 d} i \int dt' \, \delta(t'-t) H_S \frac{d}{dt'}\ket{\psi_S(t')} \nn\\
&= H_S \ket{\psi_S(t)} + \frac{G}{c^4 d} i H_S \frac{d}{dt}\ket{\psi_S(t)}, \nn
\end{align}
which upon rearranging gives
\begin{align}
i  \frac{d}{dt} \ket{\psi_S(t)}  &=  \frac{H_S}{I_S - \frac{G}{c^4 d}  H_S} \ket{\psi_S(t)} \nn \\
&=\left[ H_S +\frac{G}{c^4 d}  H_S^2  + \mathcal{O}\!\left(\tfrac{G^2}{c^8 d^2}\right)\right] \ket{\psi_S(t)}. \label{gravmodification}
\end{align}
From Eq.~\eqref{gravmodification} we see that the effect of a gravitational interaction coupling  the clock and system is such that the system Hamiltonian gets corrected at order $G/c^4$ and the strength of this correction is inversely proportional to the distance $d$ between the clock and system.}
 
\subsection{{Qubit clock and system}}

{In this example we consider both the clock and system to be two-level systems, $\mathcal{H}_C \simeq \mathbb{C}^2$ and $\mathcal{H}_S \simeq \mathbb{C}^2$, with Hamiltonians $H_C = \Omega \sigma_z$ and $H_S = \Omega \sigma_z$. As defined below Eq.~\eqref{ClockStates}, the clock state indicating the time $t$ is  $\ket{t} \ce e^{-i H_C t} \ket{t_0}$, where we choose $\ket{t_0} \ce \tfrac{1}{\sqrt{2}} \left( \ket{0} +  \ket{1} \right)$. The identity on $\mathcal{H}_C$ may be resolved in terms of these clock states as
\begin{align}
I_C = \frac{\Omega}{\pi}  \int_0^{\frac{2 \pi}{ \Omega}} dt' \ket{t'}\!\bra{t'}. \nn 
\end{align}
Suppose that the clock and system are coupled through the interaction Hamiltonian
\begin{align}
H_{int} &=  a \left(\sigma_x \otimes \sigma_x  - \sigma_y \otimes \sigma_y \right)  \nn \\
& \quad + b \left( \sigma_z \otimes \sigma_z + I_C \otimes I_S \right), \label{QubitInteraction}
\end{align}
where $a,b \in \mathbb{R}$.}

{In this case, the conditional state of the system $\ket{\psi_S(t)}$ satisfies the modified Schr\"{o}dinger equation 
\begin{align}
i \frac{d}{dt} \ket{\psi_S(t)} &= H_S \ket{\psi_S(t)} \nn \\
&\quad + \frac{\Omega}{ \pi}  \int dt'\,  K(t,t') \ket{\psi_S(t)},  \nn
\end{align}
where 
\begin{align}
K(t,t') &\ce \braket{t | H_{int} | t'} \nn \\
&= a \left( e^{-i\Omega (t+t')} \ket{0}\!\bra{1} + e^{i\Omega (t+t')} \ket{1}\!\bra{0} \right)\nn \\
&\quad + b\left( e^{i\Omega (t-t')} \ket{0}\!\bra{0} + e^{-i\Omega (t-t')} \ket{1}\!\bra{1} \right) \label{QubitK}.
\end{align}
This modified Schr\"{o}dinger equation can be solved perturbatively using Eq.~\eqref{nordersol}. The first order correction to the propagator is
\begin{align}
V_1(t,t_0) &\ce  -i   U(t,t_0) \nn \\
&\quad \times \int_{t_0}^{t} ds \, U(s,t_0)^\dagger  \int du \, {K}(s,u) U(u,t_0), \nn
\end{align}
where $U(t,t') \ce e^{-i \Omega \sigma_z (t-t')}$. Given ${K}(t,t')$ in Eq.~\eqref{QubitK} the first integrand can be evaluated and shown to vanish. Thus $V_1(t,t_0) =0$, and then by Eq.~\eqref{nordersol} the $n$th order contribution to the propagator also vanishes, \mbox{$V_n(t,t_0) =0$}. We conclude that the coupling between the clock and and system described by the interaction Hamiltonian in Eq.~\eqref{QubitInteraction} has no effect on the relational dynamics between the clock and system, and thus the conditional state $\ket{\psi_S(t)}$ also satisfies the usual Schr\"{o}dinger equation, which in this case is $ i\tfrac{d}{dt} \ket{\psi_S(t)} = \Omega \sigma_z \ket{\psi_S(t)}$.}

{In this example we have the luxury of solving the associated constraint equation, Eq.~\eqref{constraint}, for the physical state $\kket{\Psi}$ directly. With the clock, system, and interaction Hamiltonians specified above, the most general solution to Eq.~\eqref{constraint} is
\begin{align}
\kket{\Psi} &= \sqrt{2} \left( \cos \theta \ket{10} + \sin \theta e^{i \phi} \ket{01} \right), \nn
\end{align}
where $\theta \in [0,\pi]$, $\phi \in [0,2\pi)$, and the physical state $\kket{\Psi}$ is normalized according to Eq.~\eqref{normalization}. Then, the conditional state at time $t$, as defined in Eq.~\eqref{SystemState}, is
\begin{align}
\ket{\psi_S(t)} &\ce \big( \bra{t} \otimes I_S\big) \kket{\Psi} \nn \\
&=  \frac{1}{\sqrt{2}} \left( e^{i \Omega t} \ \bra{0} + e^{-i \Omega t} \bra{1} \right) \otimes I_S \nn \\
&\quad \times \sqrt{2} \left( \cos \theta \ket{10} + \sin \theta e^{i \phi} \ket{01} \right)\nn \\
&= e^{i H_S t} \ket{\psi_S(t_0)}, \label{QubitStateAtTime}
\end{align}
where $\ket{\psi_S(t_0)} \ce \cos \theta \ket{0} + \sin \theta e^{i \phi} \ket{1}$ is an arbitrary state in $\mathcal{H}_S$. Note that in this example the fact that $\ket{\psi_S(t_0)}$ can be an arbitrary state in $\mathcal{H}_S$ follows from the fact that the dimension of the null space of the total Hamiltonian $H = H_C \otimes I_S + I_C \otimes H_S +  H_{int}$ is equal to the dimension of the system Hilbert space $\mathcal{H}_S \simeq \mathbb{C}^2$, $\dim \left[ {\rm Null}\left(H\right) \right] = \dim\left[\mathcal{H}_S \right] = 2$. Equation \eqref{QubitStateAtTime} confirms the conclusion stated in the previous paragraph.}

\section{Conclusions and outlook}
\label{ch8Summary}

In this article we have generalized the conditional probability interpretation of time to account for an interaction between a clock and the system whose evolution it is tracking. This is a necessary consideration if the conditional probability interpretation is to be applied to any model of quantum gravity because gravity couples everything, including any clock and system of interest. In the case of an interaction between the clock and system, we find the conditional state of the system $\ket{\psi_S(t)}$ satisfies a modified time-nonlocal  Schr\"{o}dinger equation. A series solution in the interaction strength to this modified Schr\"{o}dinger is derived.

{In Sec. IV we gave three explicit examples of clock system interactions. In the first example we found that when the interaction Hamiltonian $H_{int}$ is of the form given in Eq.~\eqref{specialinteraction}, the modified Schr\"{o}dinger equation becomes the usual Schr\"{o}dinger equation with a time-dependent Hamiltonian dependent on $H_{int}$. In the limit when the interaction between the clock and system vanishes, $H_{int}=0$, the modified Schr\"{o}dinger equation reduces to the ordinary Schr\"{o}dinger equation. In the second example the clock and system are coupled through a Newtonian gravitational interaction. In such a case we find that the modified Schr\"{o}dinger equation reduces to the usual Schr\"{o}dinger equation with a modified system Hamiltonian. This example illustrates one way in which a gravitational interaction between the clock and system could affect the induced relational dynamics and offers a toy model in which the primary motivation for our investigation is realized (gravity couples to everything). Finally, the third example examines the situation when both the clock and system are two-level systems. We consider a clock-system interaction for which the modified Schr\"{o}dinger equation is equivalent to the usual Schr\"{o}dinger equation, demonstrating that an interaction between the clock and system doesn't necessarily modify the relational dynamics governing the conditional state.  This example also illustrates the case when the clock states are not eigenstates of an operator canonically conjugate to the clock Hamiltonian.}

As it stands, the conditional probability interpretation of time does not specify a unique choice of clock states, and thus does not address the multiple choice problem~\cite{Kuchar:2011} (although \cite{Marletto:2016} offers a possible resolution). In case of a perfect clock, as discussed below Eq.~\eqref{ClockStates}, the clock states are completely delocalized in the energy basis of the clock, that is, they are eigenstates of the operator canonically conjugate to the clock Hamiltonian. These clock states are maximally asymmetric under the action of the group generated by the clock Hamiltonian, which suggests that an appropriate figure of merit for choosing the clock states may come from the resource theory of asymmetry~{\cite{Bartlett:2007, Safranek2015}}.

Future work will focus on realizing specific examples of the developed formalism. It should be noted that although the results presented were in the context of nonrelativistic quantum mechanics, in principle, there is nothing stopping the application of the conditional probability interpretation to relativistic quantum field theory and theories of quantum gravity.

Another avenue to explore is the possibility of replacing the infinite dimensional clock Hilbert space with a finite dimensional one. The canonical commutation relations between the clock Hamiltonian and time operator in Eq.~\eqref{ClockCCR} will no longer be satisfied. However, it is still possible to define a self-adjoint time operator that  satisfies an approximate canonical commutation relation with the clock Hamiltonian {\cite{Peres:1995, Massar:2008, Woods:2018}}. It will be interesting to explore the role the dimension of the clock Hilbert space plays in a classical limit. 

Another task will be to generalize the above formalism to mixed states. This will allow for the investigation of how an interaction between the system and clock effects the fundamental decoherence mechanism discussed by Gambini  \emph{et al.} \cite{Gambini:2004a, Gambini:2004b,Gambini:2004, Gambini:2009}.

\begin{acknowledgments}
We wish to thank Felix Beaudoin, Leigh Norris, Lorenza Viola, Edward Vrscay, and the attendees at the 2017 Spacetime and Information workshop for useful discussions. We also thank Hilary Snyder for producing Figure 1. This work was supported by the Natural Sciences and Engineering Research Council of Canada and the Dartmouth College Society of Fellows.
\end{acknowledgments}

\bibliographystyle{plain}
\bibliography{QuantizingTime}

\end{document}